\documentclass{article}\usepackage[hidelinks]{hyperref}

\usepackage{arxiv}

\usepackage[table]{xcolor} 
\usepackage{colortbl}  
\usepackage[utf8]{inputenc} 
\usepackage[T1]{fontenc}    
\usepackage{hyperref}       
\usepackage{url}            
\usepackage{booktabs}       
\usepackage{amsfonts}       
\usepackage{nicefrac}       
\usepackage{microtype}      
\usepackage{lipsum}		
\usepackage{graphicx}
\usepackage[square,numbers]{natbib}
\usepackage{doi}

\usepackage{graphicx}
\usepackage[utf8]{inputenc}
\usepackage[english]{babel}
\usepackage{cancel}
\usepackage{amsmath}
\usepackage{url}
\usepackage{multirow}
\usepackage{array,multirow}
\usepackage{algorithm}
\usepackage{algorithmic}
\usepackage{graphicx}
\usepackage{verbatim}
\graphicspath{ {./figure/} }
\usepackage{babel}
\usepackage{xfrac}
\usepackage{dblfloatfix}
\usepackage{etoolbox}
\makeatletter
\patchcmd{\@makecaption}
  {\scshape}
  {}
  {}
  {}
\makeatother
\usepackage{xcolor}
\usepackage{tabularx}
\usepackage{amssymb}
\usepackage{amsthm}
\usepackage{array, boldline, makecell, booktabs}

\newtheorem{definition}{Definition}

\usepackage{multirow}
\usepackage{enumitem}
\usepackage{diagbox}
\usepackage{afterpage}

\usepackage{color,soul}
\usepackage{array,multirow}
\usepackage{caption}

\usepackage{microtype}
\usepackage{booktabs}
\usepackage[format=hang,font={small,bf}]{caption}
\usepackage[utf8]{inputenc}
\usepackage{color,soul}
\setul{0.5ex}{0.3ex}
\definecolor{Green}{rgb}{0,1,0}
\setulcolor{Green}
\usepackage[bottom]{footmisc}

\usepackage{nccmath}
\usepackage{mathtools}

\usepackage{lipsum}

\newcommand\blfootnote[1]{%
  \begingroup
  \renewcommand\thefootnote{}\footnote{#1}%
  \addtocounter{footnote}{-1}%
  \endgroup
}

\usepackage{fancyhdr}
\title{Optimizing Cyber Defense in Dynamic Active Directories through Reinforcement Learning}

\author{
    Diksha Goel\textsuperscript{1,2},
    Kristen Moore\textsuperscript{1,2},
    Mingyu Guo\textsuperscript{3}, 
    Derui Wang\textsuperscript{1,2}, 
    Minjune Kim\textsuperscript{1,2}, 
    Seyit Camtepe\textsuperscript{1,2}
}

\date{}

\begin{document}
\maketitle

\vspace{-0.5in}
\begin{center}
\textsuperscript{1}CSIRO’s Data61, Australia \\
\textsuperscript{2}Cyber Security Cooperative Research Centre (CSCRC), Australia \\
\textsuperscript{3}University of Adelaide, Australia \\
\texttt{\{diksha.goel, kristen.moore, derek.wang, minjune.kim, seyit.camtepe\}@data61.csiro.au} \\
\texttt{mingyu.guo@adelaide.edu.au}
\end{center}


\vspace{0.5in}
\begin{abstract}
This paper addresses a significant gap in Autonomous Cyber Operations (ACO) literature: the absence of effective edge-blocking ACO strategies in dynamic, real-world networks. It specifically targets the cybersecurity vulnerabilities of organizational Active Directory (AD) systems. Unlike the existing literature on edge-blocking defenses which considers AD systems as static entities, our study counters this by recognizing their dynamic nature and developing advanced edge-blocking defenses through a Stackelberg game model between attacker and defender. We devise a Reinforcement Learning (RL)-based attack strategy and an RL-assisted Evolutionary Diversity Optimization-based defense strategy, where the attacker and defender improve each other’s strategy via parallel gameplay. To address the computational challenges of training attacker-defender strategies on numerous dynamic AD graphs, we propose an RL Training Facilitator that prunes environments and neural networks to eliminate irrelevant elements, enabling efficient and scalable training for large graphs. We extensively train the attacker strategy, as a sophisticated attacker model is essential for a robust defense. Our empirical results successfully demonstrate that our proposed approach enhances defender's proficiency in hardening dynamic AD graphs while ensuring scalability for large-scale AD\footnote{This work has been supported by the Cyber Security Research Centre Limited whose activities are partially funded by the Australian Government’s Cooperative Research Centres Programme.}. \blfootnote{The manuscript has been accepted as full paper at European Symposium on Research in Computer Security (ESORICS) 2024.}

\end{abstract}

\section{Introduction}
In the rapidly evolving digital world, organizations are strengthening their cybersecurity in response to the increasing frequency and severity of cyber attacks \cite{nandi2016interdicting, ahmad2023review}. Despite these efforts, traditional security operations centre analysts often face large volumes of alerts that lead to alert fatigue and chances of critical warnings being overlooked. This has motivated research into leveraging advances in Artificial Intelligence (AI) to scale and extend the capabilities of human operators to defend networks. One such emerging direction is Autonomous Cyber Operations (ACO), which involves the development of blue team (defender) and red team (attacker) decision-making agents in adversarial scenarios. Reinforcement Learning (RL) based solutions \cite{applebaum2022bridging,nguyen2021deep} have demonstrated promising results in this domain, where the agents learn optimal cyber defense policies by exploring environmental dynamics. Several platforms, such as FARLAND \cite{molina2021network}, CybORG \cite{cage_cyborg_2022}, and CyberBattleSim \cite{msft:cyberbattlesim}, have been developed to test and validate RL-based approaches in simulated cybersecurity environments. MITRE developed the FARLAND platform, which employs generative programs to model diverse network environment distributions, facilitating the development of RL-based defense mechanisms against evolving adversarial tactics. The CybORG platform, developed by the Australian Government's Department of Defense, offers a wide range of simulation environments for Ant Colony Optimization (ACO) research. It spans scenarios from safeguarding autonomous drone networks to defending defense industry enterprises. CyberBattleSim, developed by Microsoft, simulates automated red team activities in networks, emphasizing the offensive side of cybersecurity operations. Although many studies use these platforms to advance ACO, their effectiveness is limited due to the significant difference in scale and complexity between the simulated environments and real-world networks.

Another body of work in the ACO literature \cite{guo2022practical, guo2022scalable, Goel2022defending, goel2023evolving} focuses specifically on defending \textbf{\textit{Microsoft Active Directory (AD)}}. AD serves as a primary security management tool for \textit{Windows Domain Network}, enabling administrators to manage and control access to network resources. Given the widespread
\begin{figure*}
    \centering
    \includegraphics[width=0.30\textwidth]{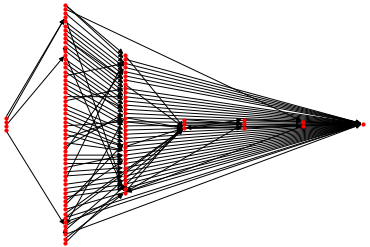} 
    \caption{AD attack graph containing 500 computers.}
    \label{fig:attack_graph}
\end{figure*} 
adoption of Microsoft Domain Network among small as well as large organizations, AD has become a prime target for cyber attackers. Reports indicate that monthly, 1.2 million Azure AD accounts are compromised, with 80\% of intrusions targeting administrative accounts \cite{microsoft2023}. These statistics highlight the importance of developing autonomous cyber defense agents to help harden AD environments. AD environments provide insights into real-world cyber defense scenarios and enable the development and validation of autonomous cyber defense
strategies. By simulating attacks and defense mechanisms within the context of AD, we can explore the complexities of real-world cyber ecosystems and develop strategies tailored to mitigate threats at a large scale, thereby enhancing the overall resilience of organizational IT infrastructures against cyber attacks.

AD graph can be represented as an attack graph, where nodes symbolize computers, accounts, or security groups, while directed edges $(i,j)$ depict trust relationships that an attacker can exploit to escalate privileges or move laterally from node $i$ to node $j$. \textsc{BloodHound} is a widely used tool to discover attack paths in AD graphs. It employs an \textit{identity snowball attack}, starting from a low-privilege account and progressing towards a high-privilege account (\(\text{Computer A} \xrightarrow{\text{HasSession}}  \text{User B} \xrightarrow{\text{AdminTo}} \text{Computer C}\)). Figure \ref{fig:attack_graph} shows an AD attack graph generated using \textsc{DBCreator}, a tool devised by the \textsc{BloodHound} team to generate synthetic AD graphs. Before BloodHound, attackers relied on trial and error in AD attack graphs to reach Domain Admin (DA). BloodHound improves this by mapping shorter, less detectable attack paths. This ease of access raises security concerns. In response, defenders start exploring strategies like edge blocking in AD attack graphs. BloodHound draws inspiration from academic research \cite{dunagan2009heat}, where researchers developed a heuristic to block edges in attack graphs. This approach aims to disconnect the graph and hinder attackers from reaching DA. In AD environments, edge blocking involves actions like access revocation or enhanced surveillance to prevent unauthorized access to DA.

Existing solutions \cite{Goel2022defending, goel2023evolving, guo2022scalable, guo2022practical, Yumeng23:Near} for hardening AD graphs overlook their dynamic nature and assume AD graphs to be static. However, real-world AD graphs constantly change primarily due to user activities like logging in and out of computers. 
\textit{\textbf{In this paper, we develop a framework for training autonomous agents to defend large-scale, \textit{dynamic} AD graphs.}} We model a Stackelberg game where the attacker aims to infiltrate and maximize their chances of reaching the highest-privilege {\textit{Domain Admin (DA)}} account in a dynamic AD graph. In contrast, the defender aims to thwart the attacker's attempts by devising an effective edge-blocking defensive policy. The dynamic nature of AD is characterized by the On/Off presence of \textsc{HasSession} edges (particular edge type in AD). These edges are added to the graph when users log into the system, and the edges remain online until the session ends. We assume each \textsc{HasSession}  edge is active with a 50\% probability. We define a \textit{graph snapshot} at time $t$ as a specific state of dynamic AD graph, with all nodes and only active \textsc{HasSession} edges at time $t$.

We address the attacker-defender problem in dynamic AD graphs by devising a \textbf{\textit{Generalized
Reinforcement Learning (GenRL) attacking policy}} and a \textbf{\textit{Reinforcement Learning-assisted Evolutionary Diversity Optimization (RL-EDO)-based defensive policy}}. The defender’s RL-EDO generates multiple diverse defenses, while the attacker’s RL agent is trained in parallel across numerous RL environments. In each environment, the attacker faces various AD graph snapshots and one of the defender's defense strategy. We train the attacker’s RL agent to optimize its success in reaching the DA in each environment. The dynamic nature of AD graphs presents a challenge due to the exponential number of potential snapshots, each representing a possible starting point for the attacker. This results in an exponential number of RL environments when we use RL to train the attacker. However, many of these environments are highly similar due to the high degree of similarity among the snapshots. This similarity allows the knowledge learned from one environment to transfer to others. To train the agent to learn a shared policy and maximize rewards, we train the RL agent across multiple environments in parallel. During this parallel training, the RL agent is exposed to numerous different snapshots in each environment, broadening its exposure to a wider range of possibilities. This helps the attacker learn generalized knowledge applicable to dynamic graph settings, enhancing its ability to navigate the complexities of dynamic AD graphs effectively.

To address the computational challenge of training the attacker policy against an exponential number of RL environments, we propose \textit{\textbf{RL Training Facilitator (TrnF)}} that performs environment and Neural Network (NN) pruning to streamline the attacker's RL training process. \textit{Environment pruning} involves simplifying the RL environment by eliminating elements irrelevant to the attacker's goals. For instance, if the attacker never utilizes certain edges,  then there is no need to track their dynamic changes; thus, we can effectively disregard them. Similarly, \textit{Neural Network pruning} optimizes NN architectures by reducing the weight of less significant dimensions. The proposed RL training facilitation technique serves to accelerate the attacker's training pace while also enhancing the performance of the RL agent. \textit{Notably, we extensively train the attacker's policy as it is essential to have a well-trained attacking policy to develop an effective defense policy. }

Existing literature consists of solutions for defending dynamic AD graphs via node-blocking strategy~\cite{ngo2024optimizing, huang2023co}. However, the research problem they have considered is different from ours as we focus on defending dynamic AD graphs via edge-blocking defense. Moreover, \cite{ngo2024optimizing, huang2023co} studied a trivial attacker model where the attacker aims to reach DA via the shortest path, and if this path does not lead directly to the DA, the attack ends. This simplistic approach makes their attacker policy easy to predict, in turn, making it easier for the defender to defend. In contrast, our model presents a challenging planning problem for both attacker and defender. In our model, the attacker encounters a novel game during training that the attacker has never experienced before. Likewise, the defender is unaware of the attacker's strategies, adding uncertainty to the defense and making our problem more difficult. Consequently, a gap exists in the literature, i.e., the absence of effective edge-blocking strategies for dynamic AD networks, and our work is the first attempt to address this issue.\\

\noindent Our main contributions are summarized as follows:

\begin{itemize}[leftmargin=*]
    \item \textbf{\textit{Attacker policy.}} We propose a Generalized RL attacking policy, trained across multiple RL environments concurrently. This approach accelerates convergence and enhances performance through shared learning experiences.
    
    \item  \textbf{\textit{Defender policy.}} We design an RL-EDO based defensive strategy that generates and optimizes defense mechanisms. Unlike traditional defenses, RL-EDO adapts to sophisticated attack strategies by dynamically replacing ineffective defenses with more robust alternatives.
    
    \item  \textbf{\textit{RL training facilitator.}} To address the scalability challenges in RL training for large AD, we design an innovative RL training facilitator. It optimizes the training process by pruning irrelevant elements from the environment and neural network architectures, ensuring efficient learning without compromising defense effectiveness.
   
    \item \textbf{\textit{Experimental analysis.}} We perform experiments on varying sizes of AD graphs, i.e., r1000\footnote{r1000 represents an AD graph containing 1000 computers.}, r2000, and r4000. Our results demonstrate that 1) Our proposed attacker-defender approach generates highly effective defense; 2) Our approach accurately models the attacker problem in dynamic AD graphs; 3) Our approach is scalable to very large-scale dynamic AD graphs.    
    \end{itemize}


\section{Related Work}
\noindent \textit{\textbf{Defending Active Directories.}} Guo et al. \cite{guo2022practical} proposed an FPT algorithm and a graph neural network-based approach for defending AD graphs. In another study, Guo et al. \cite{guo2022scalable} developed a dynamic program and an RL-based approach for hardening large AD graphs. Goel et al. \cite{Goel2022defending} introduced a neural network and EDO-based approach to address the attacker-defender problem, aiming to formulate an effective defensive policy. Goel et al. \cite{goel2023evolving} developed an RL-based attacker policy for hardening large-scale AD graphs. Guo et al. \cite{guo2024limited} investigated the optimal edge-blocking problem, focusing on strategies that require minimal human intervention. Zhang et al. \cite{Yumeng23:Near} devised a dual oracle solution for defending AD and evaluated it against industrial solutions. The aforementioned approaches are designed for static graphs and do not effectively address the challenges associated with dynamic AD graphs. Ngo et al. \cite{quang} proposed a defensive strategy for placing honeypots on network nodes to defend dynamic AD graphs. In another study, Ngo et al. \cite{ngo2024optimizing} proposed an EDO-based decoy placement solution for time-varying AD graphs. However, both studies \cite{quang, ngo2024optimizing} focused on node-blocking strategies for defending dynamic AD. Our objective is to intercept edges rather than nodes, making their solutions inapplicable to our problem.
\vspace{0.08in}

\noindent \textit{\textbf{Autonomous Cyber Defense.}} CyBORG offers simulated environments for autonomous cyber defense through its 4 CAGE challenges \cite{cage_challenge_announcement,cage_challenge_2_announcement,cage_challenge_3_announcement,cage_challenge_4_announcement}. These challenges aim to enhance the blue agent's capabilities to defend against red team attack in various scenarios, such as autonomous drone networks and adversarial cyber-physical systems. Various RL approaches \cite{foley2022autonomous,heckel2023neuroevolution,hicks2023canaries,bates2023reward,foley2022autonomous,prebot2022cognitive} have been developed to advance autonomous cyber defense abilities. However, while CyBORG is useful for exploring cyber defence, its small scale and simple structure limit its applicability to real-world networks, which are comparatively larger and more complex. Consequently, solutions designed for CyBORG may not be able to handle the scalability and complexity issues associated with AD graphs.
\vspace{0.08in}

\noindent \textit{\textbf{Evolutionary Diversity Optimization.}} Hebrard et al. \cite{hebrard2005finding} devised a strategy for discovering diverse solutions in constrained programming. Do et al. \cite{do2022analysis} examined various EDO techniques for permutation problems. Neumann et al. \cite{neumann2022evolutionary} developed EDO algorithms to address the stochastic version of the knapsack problem. Neumann et al. \cite{neumann2022coevolutionary} proposed a coevolutionary pareto diversity optimization approach for enhancing constrained single-objective problems. Nikfarjam et al. \cite{nikfarjam2023evolutionary} investigated the integration of EDO algorithms with SAT solvers to maximize diversity in heavily constrained boolean satisfiability problems.

\section{Problem Description}
We investigate a Stackelberg game involving a single attacker and a defender in a directed dynamic AD graph $G = (V, E)$, where $V$ represents the nodes and $E$ represents the edges. The node set \(V\) remains constant, while the edge set \(E\) dynamically changes due to user activities. This dynamism is primarily influenced by the presence or absence of \textsc{HasSession} edges (\(H \subseteq E\)), which are the key reasons for changes in real-world AD graphs. We assume that each \textsc{HasSession} edge is present with a 50\% probability. Let $C$ and $U$ denote the set of computers and users in the AD graph, respectively. Authentication data for modelling user activities in AD graph can be denoted as $\langle t_{\text{start}}, t_{\text{end}}, u_i, c_j \rangle$, where $t_{\text{start}}, t_{\text{end}}, u_i$, and $c_j$ represent the sign-in time, sign-off time, user, and computer, respectively. In the real-world, attackers may employ tools like \textsc{SharpHound} to extract sign-in and sign-off times from Windows logs. A {\textit{Graph Snapshot}} at time $t$ can be represented as $G_t = (V, E_t)$, where $E_t = \langle e_{0,t}, e_{1,t}, \ldots, e_{m-1,t}, e_{m,t} \rangle$, with $m$ denoting the total number of \textsc{HasSession} edges. Each $e_{i,t}$ indicates whether the \textsc{HasSession} edge is active at time $t$, with 1 representing active and 0 representing inactive. Graph snapshots are represented as $G_s = \{G_1, G_2, \ldots, G_l\}$, where $l$ denotes the possible number of snapshots. AD graph comprises $s$ \textit{entry nodes}, enabling the attacker to initiate an attack from any of these nodes, and there is a single \textit{Domain Admin} (DA). The attacker aims to devise an attacking policy to maximize their probability of reaching DA across all possible snapshots. On the other hand, the defender's goal is to minimize the attacker's success probability by selectively blocking $k$ edges, where $k$ is the defensive budget. Edge blocking in AD is a costly security measure, as it necessitates extensive auditing of access logs to remove edges safely. Consequently, budgets allocated for this process are typically low. Notably, not all edges are blockable; only specific edges labelled as `blockable' can be blocked. Each edge $e$ in the AD graph is associated with a detection probability, failure probability, and success probability. The \textit{detection probability} $p_{d(e)}$ represents the likelihood that an attacker traversing edge $e$ is detected and subsequently, the attack is terminated. \textit{Failure probability} $p_{f(e)}$ indicates the chances that an attacker fails to traverse edge $e$ for reasons such as being unable to crack a password, etc. In such instances, the attack is not terminated, allowing the attacker to explore other unexplored edges. The \textit{success probability} $p_{s(e)}$ denotes the chances of attacker successfully traversing an edge and is calculated as $(1 - p_{d(e)} - p_{f(e)})$. The attacker starts an attack from a starting node and systematically explores unexplored edges to reach the DA until detection, exhaustion of all options, or successfully accessing DA. Additionally, the attacker maintains a \textit{Checkpoint} set, which records the nodes under their control. This set serves as an alternate plan for continued attack upon failure, enhancing the attacker's strategic approach.

\section{Proposed Attacker-Defender Approach}
This section first presents our proposed AD graph optimization technique, followed by RL-based attacking policy, RL training facilitator, and RL assisted evolutionary diversity optimization-based defensive approach. Finally, we discuss our overall attacker-defender strategy.

\subsection{Proposed AD Graph Optimization Technique}
The original AD graph is highly complex, making it difficult to process in its original state. To address this, we propose a graph optimization technique that utilizes structural features to create a more condensed representation. Below are some terminologies used in creating this condensed graph.

\begin{definition}
\textit{{Splitting nodes}} are the nodes that have more than one outgoing edge. $\textsc{Split}$ denotes the set of splitting nodes.
\end{definition}

\begin{definition}
{{Entry nodes}} are the starting points from where an attacker can initiate an attack. $\textsc{Entry}$ represents the set of entry nodes.
\end{definition}

\begin{definition}
\textit{{Non-Splitting Path (NSP)}} from node $x$ to $y$ is a path that begins at node $x$ and solely reaches node $y$, where $y$ is the only successor of $x$. From node $y$, the path extends to its only successor until it reaches the DA or another splitting node \cite{guo2022practical}.
\begin{equation*}
        NSP = \{ \text{{NSP}}(x, y)\}
    \end{equation*}
    \text{where} $x \in \textsc{Split} \cup \textsc{Entry}$ and $y \in \textsc{Successors}(x)$.
\end{definition}

\begin{definition}
\textit{{Block-Worthy edge (BW)}}. A block-worthy edge $bw(x, y)$ is an edge on path $NSP(x, y)$ that can be blocked and is located farthest away from node $x$. The set of block-worthy edges is denoted as:
    \begin{equation*}
    BW = \{bw(x, y) \}
    \end{equation*}
    where $x \in \textsc{Split} \cup \textsc{Entry}$ and $y \in \textsc{Successors}(x)$.
\end{definition}
\noindent A block-worthy edge may be shared among two or more NSPs. For each NSP, we allocate single unit of budget for blocking purposes. {\textit{Our AD graph optimization approach reduces the initial  AD graph with $n$ nodes and $m$ edges to a graph containing  $(|\textsc{Entry}| + |\textsc{Split}| + 1)$ nodes and $|\text{NSP}|$ edges.}}

\subsection{Attacker Approach: Reinforcement Learning}
The attacker aims to develop an attacking strategy to optimize their chances of successfully reaching the DA in any given snapshot of AD graph. We design a \textbf{\textit{Generalized Reinforcement Learning (GenRL)}} based attacking strategy. We concurrently train the RL agent across numerous defensive plans implemented across multiple graph snapshots in separate RL training environments.
\vspace{0.08in}

\noindent \textbf{\textit{Attacker’s Environment.}} Attacker’s goal of reaching DA can be formalized as a Markov Decision Process (MDP), \( M = (S, A, R, T) \), where \( S \), \( A \), \( R \), \( T \) represents the state space, action space, reward function, and transition function, respectively. MDP serves as the attacker's environment and is described below.
\begin{itemize}[leftmargin=*]
\item  \textit{{State space (S).}} The state space represents the potential states of the attacker, where each state $s \in S$ is a vector of length $|\text{NSP}|$, and each element in the state corresponds to an NSP in AD graph. Attacker’s state  $s$ is denoted as:
\begin{equation}
\label{attacker state vector}
\text{s} = \underbrace{< F, S, \ldots, ?, S, ?, S, F >}_{\text{Length = \# NSP}} 
\end{equation}
Here, `$S$’ denotes that the attacker tried this NSP and successfully made it to the other end of NSP, `$F$’ indicates a failed attempt, and `$?$’ represents that the corresponding NSP has not been tried yet. For a given state $s$, the attacker selects an NSP marked as `$?$’, attempts to traverse it and updates its status to `$S$’ or `$F$’ according to the outcome. The process continues until the attacker reaches DA, gets detected, or exhausts all options. Throughout the attack, attacker's current state $s$ serves as a \textit{knowledge base} and provides information about NSPs under attacker’s control, failed attempt NSPs, and unexplored NSPs. In this way, our sophisticated attacker keeps track of past failed attempts and avoid wasting time on those attempts again, rendering attacker’s strategy more effective. There are two terminating states: 1) If attacker ends up reaching DA, the attack ends. 2) If attacker fails to reach DA due to detection or exhaustion of all options, the attack fails and terminates.

\item  {\textit{Action space (A).}} 
For a given state $s$, the action space $A$ represents the available actions from that state, i.e., NSPs outgoing from successful NSPs in $s$. An action $a \in A$ represents an NSP and indicates that the attacker may attempt to traverse the selected NSP to reach DA.

\item  {\textit{Reward function (R).}} Reward $r(s,a)$ for state $s$ on taking action $a$ is 1, if the attacker successfully reaches DA. Otherwise, the reward is 0.

\item  \textit{{Transition function (T).}} For any state-action pair $(s, a)$, the transition function executes action $a$ on state $s$, leading to a set of potential future states. Each potential state is linked to a transition probability, which determines the likelihood of transitioning to the specific state.   
\end{itemize}

\noindent \textit{\textbf{Training Procedure.}} We utilize the actor-critic-based Proximal Policy Optimization (PPO) algorithm to train the attacker's strategy. We chose the PPO algorithm for its actor-critic framework suited to our attacker-defender RL policy and its efficient handling of discrete action spaces crucial for our AD network simulation. The \textit{actor network} proposes actions to maximize rewards, while the \textit{critic network} assesses the attacker's success rate for each state. To ensure robust training on dynamic AD graphs, we utilize 50 graph snapshots per environment, each containing a specific defense strategy devised by the defender. The attacker's initial state is determined by implementing the defense in one of these snapshots. Subsequently, the RL agent undergoes training against this snapshot with implemented defense. After each episode, a new snapshot is selected from a pool of 50, enabling training against diverse graph scenarios. The RL agent operates concurrently across multiple environments to gather data. At each step within an episode, the agent observes a state \( s_t \), selects an action \( a_t \) using the actor network, transitions to a new state \( s_{t+1} \) based on the action, and receives a reward \( r_{t+1} \). This process continues until the agent reaches DA or is detected. The training objective is to maximize cumulative rewards and learn a shared policy adaptable across different defense strategies and varying graph configurations. Initially, games may vary across environments, but as the agent learns, it refines its policy to accommodate decreasing differences between environments.
\vspace{0.08in}

\noindent \textbf{\textit{Training Challenge.}} Training the attacker's policy for dynamic AD graphs presents a significant challenge due to the exponential number of distinct snapshots available as starting points. The number of these snapshots can grow exponentially, reaching up to \( 2^{|NSPs|} \), assuming a 50\% probability for each \textsc{HasSession} edge to be active. While not every NSP necessarily includes a \textsc{HasSession} edge, we consider the worst-case scenario where at least one \textsc{HasSession} edge is present in each NSP. However, training the RL agent against every possible \( 2^{|NSPs|} \) snapshots is impractical due to the large number of NSPs present in real-world AD graphs. To address this challenge, we propose an RL training facilitator designed to streamline and optimize the RL agent's training process.

\subsection{RL Training Facilitator: Pruning Approaches}
To address the challenge of training a generalized attacker policy across numerous graph snapshots, we propose the \textbf{\textit{RL Training Facilitator (TrnF)}} to optimize the efficiency and effectiveness of the RL agent's training. Specifically, we introduce two pruning approaches aimed at streamlining the training process by removing elements irrelevant to the attacker. This reduces computational overhead and enhances the agent's capacity to learn effectively.
\vspace{0.08in}

\noindent \textit{\textbf{Environment Pruning via Simplification Agent.}} 
We introduce a \textit{simplification agent} designed to optimize the environment by identifying and removing unnecessary NSPs (which can be labelled as noise) due to their non-utilization by attackers. This agent prunes irrelevant NSPs, thereby reducing the number of potential graph snapshots that the attacker's policy needs to learn. Initially, we train the RL agent using the attacker’s policy (Section 4.2) and subsequently deploy the simplification agent for environment pruning. The simplification agent analyzes (state, action) pairs across episodes, and if the agent consistently takes specific actions (NSPs) for certain states, then it eliminates unused NSPs. Universally irrelevant NSPs are identified using the trained RL critic network, i.e., NSPs irrelevant across all environments. From this set, a subset of NSPs is randomly selected, and if their removal does not impact the critic value, they are discarded; iterative attempts with different NSP sets are conducted if there is a change in state value. We remove irrelevant NSPs from half of the environments only in order to expose the RL agent to diverse scenarios. After the iterative process, previously removed NSPs are reintroduced to confirm their irrelevance. If the RL agent still does not utilize them, it confirms their irrelevance. This reduction in NSPs limits the starting points for the attacker’s policy learning. Blocking $x$ irrelevant NSPs reduces starting points by \(2^x\), thereby accelerating training and optimizing resource allocation.

\vspace{0.08in}

\noindent \textit{\textbf{Neural Network Pruning via Weight Reduction by Fixed Ratio.}} 
We propose a NN pruning technique to optimize NN architectures for enhancing RL agent training. This technique selectively reduces weights of less critical dimensions within the NN that correspond to less influential actions. The goal is to streamline the learning process, enabling faster convergence towards optimal weights. By ignoring unnecessary dimensions in input data, the NN reduces noise and expedites training. During training, the NN evaluates the importance of dimensions for actions at split nodes and adjusts weights accordingly. We iteratively block each dimension of a split node and monitor the critic value's stability. Stable values prompt us to prune the dimension's weight by a fixed ratio, guiding the NN towards minimizing irrelevant dimension weights. This proactive approach self-corrects weight reduction errors by adjusting weights in subsequent iterations, ensuring reliability. This method actively adjusts weights rather than relying solely on training. Similar to our environment pruning technique, we leverage domain knowledge to identify and reduce unnecessary dimensions, optimizing NN architecture for faster convergence towards optimal weights.

Our RL training facilitator provides several advantages. 1) By prioritizing important NSPs, it accelerates RL policy training and optimizes resource efficiency by focusing on relevant snapshots. 2) Removing irrelevant NSPs filters out noise, improving the accuracy of attacker behaviour modelling for precise decision-making. 3) Integrating the training facilitator enhances the scalability of attacker policies by simplifying both the environment and NN, thereby reducing complexity and enabling quicker convergence. These improvements collectively enhance the performance of the RL agent significantly.

\subsection{Defender's Approach: Reinforcement Learning Assisted Evolutionary Diversity Optimization}

To defend dynamic AD graphs, the defensive approach must minimize the attacker's success rate across all potential AD graph snapshots.\textit{ However, designing individual defensive strategies for each snapshot is impractical due to the exponential number of possible graph snapshots ($2^{|NSPs|}$)}. Therefore, our goal is to devise a generalized defensive policy that minimizes the attacker's success probability across any conceivable snapshot. To address this, we propose a \textbf{\textit{Reinforcement Learning assisted Evolutionary Diversity Optimization (RL-EDO)}} policy. Our approach involves extracting a set of static graph snapshots, denoted as $G_s$, from the dynamic AD graph and strategically blocking a subset of edges in the AD graph to minimize the attacker's average success rate across all instances in $G_s$. The RL-EDO approach uses EDO to generate multiple diverse, high-quality defenses and allows the attacker to play against these defenses across multiple environments. After training the attacker's policy, the defender employs the trained RL critic network to evaluate the attacker's performance against each defense. Defenses that are advantageous for the attacker are replaced with better alternatives, avoiding the computational efforts required to train the policy against these defenses. The defender is constrained to block a maximum of $k$ edges\footnote{Only block-worthy edges can be blocked, as not all edges are blockable.}. The defender uses the attacker's trained RL critic network as a fitness metric for assessing individual defensive strategies. The fitness of a defensive plan indicates the attacker's success rate when facing that specific defense. The defensive strategy can be depicted as follows:
\begin{equation}
\label{def_state_vec}
\text{Defense plan vector} \, = \underbrace{< B, N, \,.\, .\, .\, ,\,N, B, B>}_\text{Length of vector = \#Block-worthy edges} 
\end{equation}
Here, $B$' and $N$' represent blocked and non-blocked edges, respectively. The defender creates an initial defense population $P$, each represented as a vector of size $|BW|$ (Refer to Eq. \ref{def_state_vec}). Each coordinate in the vector is either $B$ or $N$, and the count of `$B$'s in the vector equals the defensive budget $k$. To generate offspring (new defenses), the defender performs mutation or crossover operations with a probability of $0.5$ on randomly chosen individuals from $P$. These operations ensure that the total blocked edges remain within the budget $k$. Randomness is introduced into the operations by sampling a value $x$ from a Poisson distribution with a mean of 1. The mutation and crossover operations are performed as follows:\vspace{0.06in}

\noindent \textit{\textbf{Mutation Operation.}} A randomly selected individual defense $p'$ from $P$ undergoes mutation by swapping $x$ occurrences of $N$'s with $B$'s and $x$ occurrences of $B$'s with $N$'s to generate new offspring.

\vspace{0.06in}

\noindent \textit{\textbf{Crossover Operation.}} Two individuals, $p'$ and $p''$, are randomly selected from $P$. We then identify $x$ coordinates where $p'$ has $N$'s and $p''$ has $B$'s. We swap the values at these coordinates, replacing $N$'s in $p'$ with $B$'s and $B$'s in $p''$ with $N$'s. Similarly, we identify another set of $x$ coordinates where $p'$ has $B$'s and $p''$ has $N$'s and perform the swap again, substituting $B$'s with $N$'s and vice versa.
\vspace{0.06in}

\noindent \textbf{\textit{Diversity Optimization in Population.}} After generating offspring, we evaluate its fitness score and selectively incorporate it into $P$ only if its fitness falls within $(BEST \pm 0.1)$. If the offspring fails to meet this criterion, we reject it, even if it may bring potential diversity benefits to the population. This selective process balances the introduction of new genetic defense while maintaining the population's superior fitness. Our goal upon adding an individual to $P$ is to optimize population diversity by removing the individual that contributes the least to diversity. \textit{We define diversity as blocking all edges deemed block-worthy, with the objective of enhancing the diversity of blocked edges across the defense plan population.} This metric calculates the frequency with which each block-worthy edge is blocked across the population and aims to achieve an even distribution. In population $P$ of $\mu$ individuals, each individual $p_i$ can be represented as follows:

\begin{equation*}
p_i = \big((B/N, bw_1), (B/N, bw_2), ..., (B/N,bw_{|BW|})\big)
\end{equation*}
Here, `$B$' denotes the blocked status of the block-worthy edge, `$N$' denotes the non-blocked status, and $i \in {1, . . .,\mu}$. We compute the count of individuals who have blocked each block-worthy edge $bw_j$, where $j \in {1, ..., |BW|}$. This count is represented by the vector $C(bw)$ and is calculated as:
\begin{equation*}
C(bw) = (c(bw_1), c(bw_2), ..., c(bw_{|BW|}))
\end{equation*}
Here, $c(bw_1)$ represents the count of individuals out of $\mu$ that have blocked the $bw_1$ edge. The diversity of $P$ without including individual $p_i$ is represented by vector $D(C(bw)\backslash{p_i})$, and is calculated as:
\begin{equation*}
\label{diversity_eq}
D(C(bw)\backslash{p_i}) = C(bw) - p_i
\end{equation*}
The defender aims to maximize the diversity of blocked edges and compute $SortedD(C(bw)\backslash{p_i})$ as:
\begin{flalign*}
\text{\textit{SortedD}} (C(bw)\backslash{p_i}) = \text{sort}\Big(D(C(bw)\backslash{p_i})\Big)
\end{flalign*}
To optimize the population diversity, we minimize $SortedD(C(bw)\backslash{p_i})$ in descending lexicographic order. We achieve this by eliminating the individual $h$ with the lowest $SortedD(C(bw)\backslash{p_h})$ score to maximize population's diversity. If removing individual $h$ increases diversity and its fitness value is far from the best, we remove it from the population. Conversely, if newly generated offspring attain the best fitness value, they are included in $P$ regardless of their diversity, while the individual with the worst fitness value is eliminated. This approach enables the defender to create a diverse yet high-quality blocking plan.

\begin{figure*}[t!]
\centering
\includegraphics[width=0.6\paperwidth]{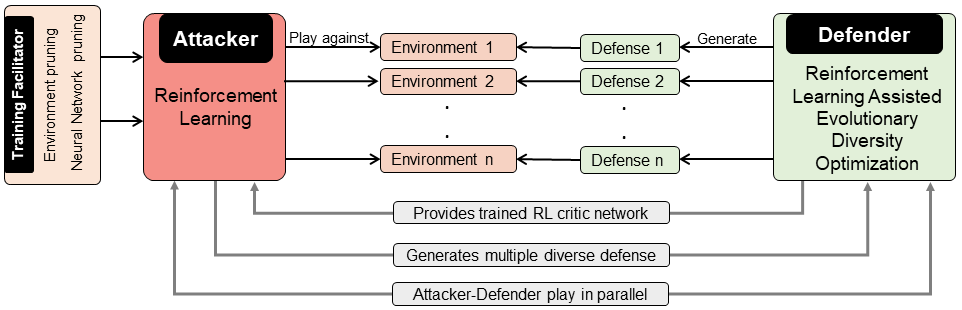}
\caption{Proposed RL-based Attacker-Defender Approach for Dynamic Networks.}
\label{architecture}
\end{figure*}

\subsection{Overall Attacker-Defender Approach}
The defender uses RL-EDO to create multiple diverse defense plans for each of the attacker's RL environments. Each RL environment contains a defense from the defender and multiple graph snapshots. Our RL agent undergoes parallel training across these RL environments, with each environment containing a defense plan from the defender implemented in numerous graph snapshots. We adopt a rotation mechanism that switches the graph snapshot the RL agent faces after each training episode. Our goal is to train a generalized attacking policy capable of achieving high success rates regardless of the AD snapshots. To address the computational challenge of training the attacker policy against numerous snapshots, we design an RL training facilitator that performs environment and NN pruning to simplify the RL training process. In environment pruning, we eliminate universally irrelevant NSPs, and in NN pruning, we reduce the weights of less important dimensions. We first train the RL agent to learn the optimal actions, and once it is trained, we perform the pruning steps. After pruning, we retrain the RL agent to optimize actions on the reduced graph. This iterative process of pruning and training ensures more efficient learning across multiple snapshots. After this iterative process, we reintroduce previously removed NSPs to verify their relevance by retraining the RL agent on the updated graph. The defender continuously evaluates the current set of defensive plans, replacing those that are advantageous for the attacker with superior alternatives. Notably, the attacker's RL critic network is utilized by the attacker's policy for implementing pruning techniques and the defender’s policy for assessing defensive strategies.
Additionally, the diversity factor helps the RL agent avoid local optima and facilitates more precise learning of the attacking policy. Overall, the parallel gameplay between the attacker and defender helps each other's policies to improve. Figure \ref{architecture} illustrates our overall proposed approach.

\section{Experimental Results}\label{Sec:Results}
We evaluate the effectiveness of our proposed attacker-defender strategy on synthetic AD graphs of varying sizes. We conduct the experiments on a high-performance cluster server, dedicating one CPU and 20 cores per trial. We perform experiments on five distinct AD graphs (using 5 different seeds ranging from 0 to 4, reporting an average over five seeds), each with varying blockable edges and entry nodes. We implemented the code in PyTorch.

\subsection{Synthetic AD Graph Dataset}\label{dataset}
In this study, we employ synthetic AD graph datasets to evaluate the effectiveness of our proposed attacker-defender strategies. Given the sensitive nature and limited accessibility of real-world AD data, synthetic datasets provide a crucial advantage in enabling controlled experiments and systematic exploration of cybersecurity strategies. We utilize \textsc{DBCreator} tool, which is a popular tool to generate synthetic AD graphs of three different sizes: r1000, r2000, and r4000, containing 1000, 2000, and 4000 computers in the graph, respectively. Details of the dataset are provided in Table \ref{dataset}. We consider three primary edge types present in \textsc{BloodHound}: \textsc{AdminTo}, \textsc{HasSession}, and \textsc{MemberOf}. For simulating attacker behaviour, we randomly select 20 starting nodes from a set of 40 nodes located at the maximum distance from the DA, ensuring coverage across a wide spectrum of potential attack paths. The probability of blocking an edge is determined based on its distance from the DA, i.e., edges farther away are more likely to be blocked. This probability is calculated as the ratio of minimum \#hops between $e$ and DA to the maximum \#hops between any edge and DA. We preprocess the AD graph by consolidating multiple DA nodes into one, removing all incoming edges to starting nodes and outgoing edges from the DA node alongside the inaccessible nodes. Each NSP is treated as a single edge.\vspace{0.08in}

\begin{table*}[t!]
\caption{Description of AD dataset.}
\label{dataset}
\renewcommand{\arraystretch}{1.3}
\centering 
\begin{tabular}{p{2cm}p{1.5cm}p{1cm}} \hline 
\textbf{{AD graph} }& \textbf{{|V|}} & \textbf{{|E|}}\\ \hline
r1000 & 2996 & 8814\\
r2000 & 5997 & 18795\\
r4000 & 12001 & 45780\\\hline
\end{tabular} 
\end{table*}

\noindent \textbf{\textit{Correlation Distributions.}} We analyze the impact of the correlation between each edge’s detection probability (\( p_{d(e)} \)) and failure probability (\( p_{f(e)} \)) on the attacker's success probability under three distributions: Independent (Ind), Positive correlation (Pos), and Negative correlation (Neg). In the Independent distribution, \( p_{d(e)} \) and \( p_{f(e)} \) are uniformly distributed between 0 and 0.2. In the Positive correlation distribution, \( p_{d(e)} \) and \( p_{f(e)} \) follow a multivariate normal distribution with mean \( \mu = [0.1, 0.1] \) and covariance matrix $\Sigma = \left[ \left[0.052, 0.5 \times 0.052\right], \left[0.5 \times 0.052, 0.052\right] \right]$. For the Negative correlation distribution, \( p_{d(e)} \) and \( p_{f(e)} \) follow a multivariate normal distribution with mean \( \mu = [0.1, 0.1] \) and covariance matrix $\Sigma = \left[ \left[0.052, -0.5 \times 0.052\right], \left[-0.5 \times 0.052, 0.052\right] \right]$.

\subsection{Training Parameters} 
For defender, the edge-blocking budget is set at 5. The defender generates a population of 20 blocking plans over 20,000 iterations\footnote{Edge-blocking is expensive due to the need to securely audit access logs for edge deletion; therefore, the budget is generally low.}. We set the crossover and mutation probabilities to 0.5. For training the attacker’s policy, we implement RL environments using OpenAI Gym \cite{brockman2016openai} and employ the PPO algorithm for training the RL agent. We implement the actor and critic networks using multi-layer NN. The model is optimized with an Adam optimizer using a learning rate of 0.0005, a batch size of 800 states, and a hidden layer size of 128. For PPO-specific hyperparameters, we follow the standard settings from the original paper \cite{schulman2017proximal}. We train the RL policy concurrently across 20 RL environments. Upon reaching the termination criterion, defender selects the defense with the least attacker success rate as the \textit{{Best Defense}}. To simulate the dynamic behaviour of AD graph, each \textsc{HasSession} edge is randomly added to graph with a probability of 0.5.

\subsection{Attacker-Defender Policy Training}\label{policy-training}
We train the attacker-defender approach for 1200 epochs, spanning approximately 4-5 days to complete the training process. During this process, the defender generates 20 diverse defensive plans, against which the attacker plays concurrently. Each RL environment contains 50 graph snapshots and a specific defense from defender. The attacker's policy undergoes continuous training, while the defender evaluates and resets defensive environments after every 50 epochs. To facilitate the training of RL agent across multiple graph snapshots, we generate 50 different graph snapshots for each environment and load a new snapshot from the pool of 50 to train the policy against diverse scenarios across all 20 environments. Upon reaching the termination condition, the defender evaluates the performance of 20 defensive plans using the RL critic network, selecting the best plan based on the lowest attacker success rate. For our proposed training facilitator, within every 50 epochs before the defender evaluates and resets the environments, we perform environment pruning in the 30\textsuperscript{th} and 40\textsuperscript{th} epochs. In the 45\textsuperscript{th} epoch, we reintroduce all removed edges, and training continues for 5 more epochs to confirm the irrelevance of the removed edges. Furthermore, we conduct environment pruning in 10 environments to expose the RL agent to both pruned and original environments. Additionally, our NN pruning technique gradually reduces the weights of less important dimensions by 2\% after every 10 minutes.

\begin{enumerate}[leftmargin=*]
    \item \textbf{GenRL-TrnF+RL-EDO (Proposed).}  GenRL is employed as attacker's policy, utilizing a training facilitator to support RL agent's training process. Meanwhile, defender employs RL-EDO approach to generate defense strategies.
    
    \item \textbf{GenRL+C-EDO  \cite{goel2023evolving}.} GenRL is employed as the attacker’s policy, while EDO is utilized as the defender’s policy. Notably, the attacker GenRL policy in this approach operates without the support of the training facilitator. 
\end{enumerate}

\subsection{Evaluating Attacker's Policy}
In this setup, our objective is to evaluate the performance of our proposed generalized attacking policy, GenRL-TrnF, and assess the impact of our training facilitator on the RL agent's learning capacity for dynamic AD graphs.\\

\begin{table*}[t!] 
\caption{Comparison of various attacker policies with 50 specialized trained RL agents (Attacker’s values closer to 50 SpecRL Agents indicate superior policy performance).}
\label{setup1}
\renewcommand{\arraystretch}{1.3}
\centering 
\begin{tabular}{p{1.2cm}p{3.4cm}p{1cm}p{1cm}p{1cm}p{1cm}p{1cm}p{1cm}p{1cm}} \hline
\multicolumn{1}{c}{\textbf{{}}}& \multicolumn{1}{c}{\textbf{{}}}&  \multicolumn{4}{c}{\textbf{{Attacker Success Rate}}} & \multicolumn{3}{c}{\textbf{{Time (hour)}}} \\  \cmidrule(lr){3-6}  \cmidrule(lr){7-9}

\textbf{Graph} & \textbf{Attacker policy} & \textbf{Ind} & \textbf{Pos} & \textbf{Neg} &  \textbf{Avg} & \textbf{Ind} & \textbf{Pos} & \textbf{Neg}\\ \hline 

\rowcolor{gray!25} & 50 SpecRL Agents &  51.83 & 53.76 & 51.76 & 52.45 & 63.73 & 60.02 & 58.09\\
r1000 & GenRL-TrnF (Proposed) & \textbf{54.69} & \textbf{49.63} & \textbf{53.31} & \textbf{52.54} & 18.56 & 19.29 & 21.33\\
&  GenRL & 46.27  & 45.94 & 45.97 & 46.06 & 15.24 & 13.51 & 16.26\\\hline

\rowcolor{gray!25} & 50 SpecRL Agents & 42.31 & 45.52 & 39.97  &  42.60 & 71.54 &   64.68 & 63.59   \\
r2000 & GenRL-TrnF (Proposed) & \textbf{40.45} & \textbf{41.91} & \textbf{42.62} & \textbf{41.66} & 23.81 & 25.55 & 26.34\\
 & GenRL & 35.89 & 35.74 & 36.55 & 36.06 & 18.49 & 19.92 & 18.85\\
\hline 

\rowcolor{gray!25} &  50 SpecRL Agents &  29.04  & 31.37& 29.14 & 29.85 & 78.25 & 73.91 & 75.03 \\
      & GenRL-TrnF (Proposed) & \textbf{32.51} & \textbf{25.83} & \textbf{25.95} & \textbf{28.09} & 26.02 & 28.43 & 28.29\\
r4000 & GenRL & 23.48 & 20.29 & 18.78 & 20.85 & 23.66 & 22.72 & 23.37\\
\hline 
\end{tabular} 
\end{table*}

\noindent \textbf{\textit{Baselines.}} The comparative attacker's policies are:
\begin{itemize}[leftmargin=*]
\item \textbf{GenRL-TrnF(Proposed).} A single generalized RL agent serves as the attacker's policy, trained to adapt to 50 distinct graph snapshots with the support of a training facilitator to enhance its training process.

\item \textbf{GenRL.} A single generalized RL agent serves as attacker's policy, learning from 50 graph snapshots independently, without using any training facilitator.
\item \textbf{50 SpecRL Agents.} RL is employed as attacker's policy without a training facilitator. Instead of using a single generalized agent, 50 distinct RL agents are trained, each dedicated to a specific snapshot. This approach aims to develop a more sophisticated attack strategy tailored to diverse scenarios.
\end{itemize}

\begin{figure*}[t!]
\centering
\includegraphics[width=6in, height=2in]{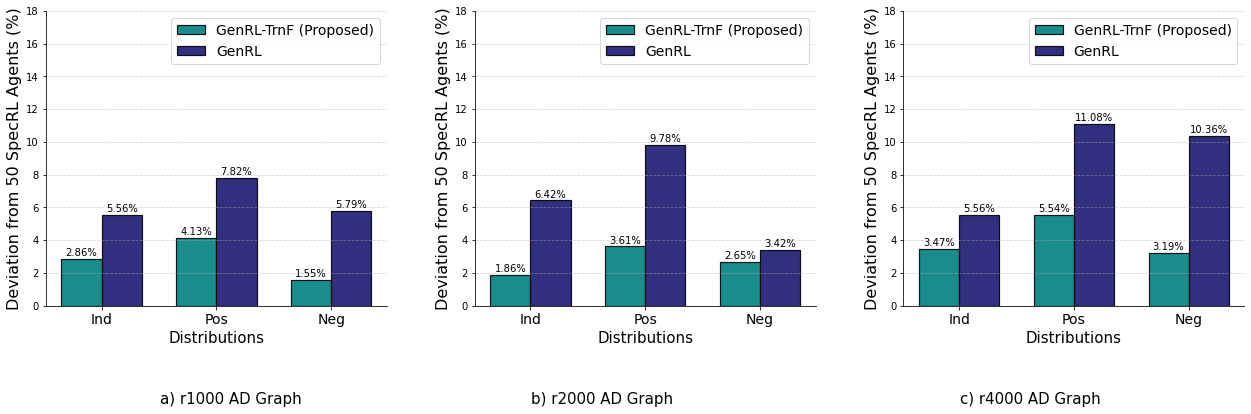}
\caption{Comparison of deviation from 50 specialized agents across various attacker policies (smaller deviations indicate superior performance). }
\label{errorplot}
\end{figure*}

\noindent \textit{\textbf{Results.}} 
To assess the performance of the attacker's policy, we deploy the best defense derived from our GenRL-TrnF+RL-EDO approach across 50 random graph snapshots. Both GenRL-TrnF and GenRL attacking policies are trained on these snapshots against the best defense for 200 epochs, and we evaluate the performance over 5000 episodes to measure effectiveness. GenRL-TrnF performs environment pruning every 30 epochs and NN pruning of 2\% every 5 minutes. We compare a single generalized RL agent trained across all snapshots against 50 specialized RL agents (50 SpecRL Agents), aiming to quantify performance differences and identify the best attacker strategies. Results averaged over five seeds (0 to 4) of AD graphs are presented in Table \ref{setup1}. Our proposed GenRL-TrnF consistently outperforms the GenRL attacking policy and closely matches the performance of 50 SpecRL Agents across all graph scales. For example, on the r1000 AD graph (Ind distribution), GenRL-TrnF achieves a 54.69\% success rate, deviating by only 2.86\% from 50 SpecRL Agents, while GenRL achieves 46.27\%, deviating notably by 5.56\%. Similarly, on the r2000 AD graph (Ind distribution), GenRL-TrnF achieves a 40.45\% success rate with a smaller deviation of 1.86\%, compared to GenRL's 35.89\% success rate with a deviation of 6.42\% from 50 SpecRL Agents. The deviation of GenRL-TrnF and GenRL from 50 SpecRL Agents is illustrated in Figure \ref{errorplot}. Our findings demonstrate that integrating a training facilitator into a generalized attacker policy enables GenRL-TrnF to perform competitively with 50 specialized RL agents, showing only slight deviations in success rate. Conversely, GenRL struggles to generalize effectively across attacker problem accurately, underscoring the crucial role of the training facilitator in enhancing RL policy efficacy by accurately modelling dynamic attacker behaviours.

\subsection{Evaluating Defender's Policy}
This section assesses the performance of our proposed defensive strategy.
\vspace{0.06in}

\noindent \textit{\textbf{Baseline.}} We compare the best defense from our GenRL-TrnF+RL-EDO approach with the GenRL+C-EDO approach \cite{goel2023evolving}. In the GenRL-TrnF+RL-EDO approach, GenRL serves as the attacker's policy with the support of a training facilitator, while the defender utilizes RL-EDO. In contrast, the GenRL+C-EDO approach employs RL alone for the attacker without any training facilitator, coupled with C-EDO for the defender.

\vspace{0.06in}

\noindent \textit{\textbf{Results.}} 
We train both approaches, GenRL-TrnF+RL-EDO and GenRL+C-EDO, using the attacker-defender approach discussed in section \ref{policy-training} to obtain the best defense. Subsequently, we generate 50 random AD graph snapshots. For each snapshot, we train one specialized RL agent integrated with the training facilitator (GenRL-TrnF) to play against the best defense obtained. We train each specialized RL agent for 200 epochs and we evaluate the GenRL-TrnF policy's performance against the best defense through simulations over 5000 episodes. The average success rate across 50 trained agents is reported in Table \ref{setup2}. The defense yielding the minimal attacker success rate is identified as the best-generalized defense. Our results consistently show that the defense from GenRL-TrnF+RL-EDO outperforms the defense from GenRL+C-EDO in reducing the attacker's success rates across all graph instances. For instance, on the r2000 graph (Ind distribution), the attacker success rate against the GenRL-TrnF+RL-EDO defense is 42.37\%, lower than the 46.05\% success rate against the GenRL+C-EDO defense. Similarly, for r1000 and r4000 AD graphs, the defense from the GenRL-TrnF+RL-EDO approach effectively reduces the attacker's success rate compared to the GenRL+C-EDO approach. Our results demonstrate that the proposed GenRL-TrnF+RL-EDO approach consistently generates superior defense against dynamic AD graphs compared to the baseline approach\footnote{The dynamic nature of AD graph problem poses significant challenges to achieving substantial reductions in attacker success rates. Even marginal decreases in success rates can yield significant benefits, considering the substantial costs associated with security breaches for organizations.}.

\begin{table*}[b!] 
\caption{Comparative analysis of best defense from various attacker-defender approaches (smaller values indicate superior performance).}
\label{setup2}
\renewcommand{\arraystretch}{1.3}
\centering 
\begin{tabular}{p{1.5cm}p{5cm}p{1.5cm}p{1.5cm}p{1.5cm}p{1cm}} \hline
\multicolumn{1}{c}{\textbf{{}}}& \multicolumn{1}{c}{\textbf{{}}}&  \multicolumn{4}{c}{\textbf{{Attacker Success Rate}}} \\  \cmidrule(lr){3-6} 

\textbf{Graph} & \textbf{Best defense from Policy} & \textbf{Ind} & \textbf{Pos} & \textbf{Neg} &  \textbf{Avg} \\ \hline 
      & GenRL-TrnF+RL-EDO (Proposed) & \textbf{56.24}  & \textbf{51.02} & \textbf{55.97} & \textbf{54.41}\\
r1000 & GenRL+C-EDO & 59.03 & 56.13 & 57.21 & 57.45\\\hline 
      & GenRL-TrnF+RL-EDO (Proposed)  & \textbf{42.37} & \textbf{42.43} & \textbf{44.65} & \textbf{43.15}\\
r2000 &  GenRL+C-EDO & 46.05 & 43.51 & 45.80 & 45.12\\\hline 
      & GenRL-TrnF+RL-EDO (Proposed) & \textbf{33.96} & \textbf{27.45} & \textbf{27.01} & \textbf{29.47}\\
r4000 & GenRL+C-EDO & 35.72  & 28.59  & 28.43 & 30.91\\\hline 
\end{tabular} 
\end{table*}

\subsection{Discussion}
Our empirical findings underscore the superior performance of our GenRL-TrnF attacker policy compared to baseline approaches. This improvement is primarily attributed to our innovative training facilitator, which enhances the efficiency of attacker training by systematically pruning irrelevant elements, guided by extensively trained RL agents. We further validated the irrelevance of these elements using a trained RL critic network, ensuring that the critic value before and after removal remains the same. As a result, our generalized GenRL-TrnF attacker policy achieves performance levels comparable to specialized agents without compromising critical network dynamics. By focusing on relevant elements identified through RL agent training, we reduce computational load and strengthen learning capacity. Furthermore, our defensive strategy significantly reduces the attacker's success rate through extensive training augmented by the training facilitator. Our integrated attacker-defender approaches reinforce each other, where a more robust attacking policy contributes to a resilient defense. Concurrent training of the RL attacker policy across multiple environments enables quicker learning of shared policies. Our proposed defense strategy demonstrates versatility in enhancing network security across various sectors: enterprise networks prevent unauthorized lateral movement, cloud environments safeguard resources and data integrity, IoT mitigates cyber-physical risks, and critical sectors like utilities, healthcare, and financial systems ensure operational resilience. Our model currently includes three edge types: AdminTo, HasSession, and MemberOf. However, real-world AD environments feature a broader range of edge types, which limits our ability to fully capture their complexities and vulnerabilities. Our future research aims to expand the model to encompass additional edge types, thereby enhancing the accuracy of our simulations and defense strategies for AD environments. Although synthetic AD graphs effectively simulate key aspects of real-world environments, they have inherent limitations in replicating complex dynamics and vulnerabilities. Validation against real AD datasets is essential to ensure the generalizability of our findings to practical cybersecurity scenarios. In future research, we will focus on validating our results using real-world AD datasets. 

\section{Conclusion}
In this study, we proposed a dual RL-based strategy for both attacker and defender within dynamic AD graphs. Our innovative training facilitator simplifies the AD graph and neural network structures, enhancing the overall efficacy of our training policy and ensuring scalability to large AD graphs. We conducted experiments on dynamic AD graphs of three different scales: r1000, r2000, and r4000. The empirical evidence demonstrates the superior performance of our approach compared to the baseline, significantly improving both the attacker's and defender's performance in dynamic network settings.

\bibliographystyle{unsrtnat}
\bibliography{references}

\end{document}